\documentclass[a4paper,10pt]{llncs}

\usepackage[utf8x]{inputenc}
\usepackage{verbatim}
\usepackage{enumerate}
\usepackage{amssymb}
\usepackage{todonotes}
\usepackage{algpseudocode}
\usepackage{graphicx}
\usepackage{textcomp}
\usepackage{algorithm}
\usepackage{amsmath}

\title{Space Efficient Algorithms for Breadth-Depth Search}
\author{Sankardeep Chakraborty\inst{1}\thanks{This work was partially supported by JST CREST Grant Number JPMJCR1402.}, Anish Mukherjee\inst{2}, Srinivasa Rao Satti\inst{3}}
\institute{RIKEN Center for Advanced Intelligence Project,\\
1-4-1 Nihonbashi, Chuo-ku, Tokyo, Japan\\
\email {sankar.chakraborty@riken.jp}\\
\and 
Chennai Mathematical Institute,\\
H1 SIPCOT IT Park, Siruseri, Chennai, India\\
\email {anish@cmi.ac.in}\\
\and
Seoul National University, \\
1 Gwanak-ro, Gwanak-gu, Seoul, South Korea\\
 \email {ssrao@cse.snu.ac.kr}\\
}
\date{}

\begin{document} \maketitle
\begin{abstract}
Continuing the recent trend, in this article we design several space-efficient algorithms for two well-known graph search methods. Both these search methods share the same name {\it breadth-depth search} (henceforth {\sf BDS}), although they work entirely in different fashion. The classical implementation for these graph search methods takes $O(m+n)$ time and $O(n \lg n)$ bits of space in the standard word RAM model (with word size being $\Theta(\lg n)$ bits), where $m$ and $n$ denotes the number of edges and vertices of the input graph respectively. Our goal here is to beat the space bound of the classical implementations, and design $o(n \lg n)$ space algorithms for these search methods by paying little to no penalty in the running time. Note that our space bounds (i.e., with $o(n \lg n)$ bits of space) do not even allow us to explicitly store the required information to implement the classical algorithms, yet our algorithms visits and reports all the vertices of the input graph in correct order.
\end{abstract}

\section{Introduction}
Graph searching is an efficient and widely used bookkeeping method for exploring the vertices and edges of a graph. Given a graph $G$, a typical graph search method starts with an arbitrary vertex $v$ in $G$, marks $v$ as visited, and systematically explores other unvisited vertices of $G$ by iteratively traversing the edges incident with a previously visited vertex. The ordering in which the next vertex is chosen from an already visited vertex yields different vertex orderings of the graph. Two of the most popular and widely used graph search methods are {\it depth-first search} ({\sf DFS}) and {\it breadth-first search} ({\sf BFS}). {\sf BFS} tries to explore an untraversed edge incident with the least recently visited vertex, whereas {\sf DFS} tries to explore an untraversed edge with the most recently visited vertex. Both of these search methods have been successfully employed as backbones for designing other powerful and efficient graph algorithms. Researchers have also devised other graph search methods~\cite{CorneilK08}, and explored their properties to design efficient graph algorithms. For example, Tarjan and Yannakakis~\cite{TarjanY84} introduced a graph search method, called {\it maximum cardinality search} ({\sf MCS}) and used it to design a linear time algorithm for chordal graph recognition and other related problems. 

Our focus here is to study another graph search method, namely {\it breadth-depth search} ({\sf BDS}) from the point of view of making it space efficient. We note that, two very different graph search strategies exist in the literature, but surprisingly, under the same name. Historically, Horowitz and Sahni~\cite{HorowitzS78}, in 1984, defined {\sf BDS} and demonstrated its applications to branch-and-bound strategies. Henceforth we will refer to this version of {\sf BDS} as $BDS_{hs}$ after Horowitz and Sahni. Greenlaw, in his 1993 paper~\cite{Greenlaw93}, proved that $BDS_{hs}$ is inherently sequential by showing it is {\sf P-complete}. Almost a decade later, Jiang~\cite{Jiang93}, in 1993, defined another graph search method, under same name {\sf BDS}, while designing an I/O- and CPU-optimal algorithm for decomposing a directed graph into its strongly connected components (SCC). In particular, he devised and used {\sf BDS} (note that, this is different from $BDS_{hs}$~\cite{HorowitzS78} as we will see shortly) to give an alternate algorithm for SCC recognition. We will refer to this version of {\sf BDS} as $BDS_{j}$ after Jiang. Implementing either of these algorithms takes $O(m+n)$ time and $O(n \lg n)$ bits of space in the standard word RAM model, where $m$ and $n$ denotes the number of edges and vertices of the input graph respectively. Our goal in this paper is to improve the space bound of the classical implementations without sacrificing too much in the running time.

\subsection{Motivation and Related Work}
Recently, designing space efficient algorithms has become enormously important due to their applications in the presence of fast growth of ``big data" and the escalation of specialized handheld mobile devices and embedded systems that have a limited supply of memory i.e., devices like Rasberry Pi which has a huge use case in IoT related applications. Even if these mobile devices and embedded systems are designed with large supply of memory, it might be useful to restrict the number of write operations. For example, on flash memory, writing is a costly operation in terms of speed, and it also reduces the reliability and longevity of the memory. Keeping all these constraints in mind, it makes sense to consider algorithms that do not modify the input and use only a limited amount of work space. One computational model that has been proposed in algorithmic literature to study space efficient algorithms, is the read-only memory ({\sf ROM}) model. Here we focus on space efficient implementations of {\sf BDS} in such settings. 

Starting with the paper of Asano et al.~\cite{AsanoIKKOOSTU14} who showed how one can implement {\sf DFS} using $O(n)$ bits in ROM, improving on the naive $O(n \lg n)$-bit implementation, the recent series of papers~\cite{Banerjee2018,Cha_thesis,Chakraborty00S18,CRS17,Chakraborty2018,ElmasryHK15}
presented such space-efficient algorithms for a variety of other basic and fundamental graph problems: namely {\sf BFS}, maximum cardinality search, topological sort, connected components, minimum spanning tree, shortest path, dynamic {\sf DFS}, recognition of outerplanar graph and chordal graphs among others. We add to this small yet rapidly growing body of space-efficient algorithm design literature by providing such algorithms for both the {\sf BDS} algorithms, $BDS_{hs}$ and $BDS_{j}$. In this process, we also want to draw attention to the fact that, even though these two search methods have same name, they work essentially in different manner. To the best of our knowledge, surprisingly this fact does not seem to be mentioned anywhere in the literature. 

We conclude this section by briefly mentioning some very recent works on designing space efficient algorithms for various other algorithmic problems: Longest increasing subsequence~\cite{KiyomiOOST18}, geometric computation~\cite{BanyassadyKMRRS17} among many others.

\subsection{Model of Computation and Input Representation} 
\label{model1}
Like all the recent research that focused on designing space-efficient graph algorithms (as in~\cite{AsanoIKKOOSTU14,Banerjee2018,Chakraborty00S18,CRS17,Chakraborty2018,KiyomiOOST18,LincolnWWW16}), here also we assume the standard word RAM model for the working memory with words size $w = \Theta (\lg n)$ bits where constant time operations can be supported on $\Theta(\lg n)$-bit words, and the input graph $G$ is given in a read-only memory with a limited read-write working memory, and write-only output. We count space in terms of the number of bits in workspace used by the algorithms. Throughout this paper, let $G=(V,E)$ denote a graph on $n=|V|$ vertices and $m=|E|$ edges where $V =\{v_1, v_2, \cdots, v_n\}$. We also assume that $G$ is given in an adjacency array representation, i.e., an array of length $|V|$ where the $i$-th entry stores a pointer to an array that stores all the neighbors of the $i$-th vertex. For the directed graphs, we assume that the input representation has both in/out adjacency array for all the vertices.

\subsection{Our Main Results and Organization of the Paper }
We start off by introducing $BDS_j$ and $BDS_{hs}$ in Sections~\ref{BDS1} and~\ref{BDS2} respectively as defined in~\cite{HorowitzS78} and~\cite{Jiang93} along with presenting their various space efficient implementations before concluding  in Section~\ref{conclusion} with some concluding remarks and future directions. Our main results can be compactly summarized as follows.

\begin{theorem}\label{combined}
Given a graph $G$ with $n$ vertices and $m$ edges, the $BDS_j$ and $BDS_{hs}$ traversals of $G$ can be performed in randomized $O(m \lg^* n)$ time\footnote{We use $\lg$ to denote logarithm to the base $2$.} using $O(n)$ bits with high probabality ($(1-1/n^c)$, for some fixed constant $c$), or
$O(m+n)$ time using $O(n \lg (m/n))$ bits, respectively.
\end{theorem}

\subsection{Preliminaries}
We use the following theorem repeatedly in our algorithms.

\begin{theorem}\cite{DemaineHPP06}\label{dict}
Given a universe $U$ of size $u$, there exists a dynamic dictionary data structure storing a subset $S \subseteq U$ of cardinality at most $n$ using space $n \lg (u/n)+ nr$ bits where $r \in O(\lg n)$ denotes the size of the satellite data attached with elements of $U$. This data structure can support membership, retrieval (of the satellite data), insertion, and deletion of any element along with its satellite data in $O(1)$ time with probabality $(1-1/n^c)$, for some fixed constant $c$. 
\end{theorem}

\section{Breadth-depth Search of Jiang}\label{BDS1}
A $BDS_{j}$ traversal of a graph $G$ walks through the vertices of $G$ and processes each vertex one by one according to the following rule. Suppose the most recently traversed edge is $(u,w)$. If $w$ still has an unvisited edge, then select this edge to traverse. Otherwise choose an unvisited edge incident on the node most recently visited that still has unvisited edges. At this point (see line $7$ of the pseudocode for $BDS_{j}$ provided below) we also say that the node $w$ is being expanded. Note that a vertex $v$ might be visited many times via different edges and here we are only interested in the last visit to the vertex (in contrast to the {\sf BFS} and {\sf DFS} where only the first visit to the vertex is considered) when $v$ is expanded. 

To implement this, more specifically to capture the fact of last visit, Jiang used an \emph{adaptive stack} (abbreviated as {\it adp-stack} in the pseudocode below). A stack is called adaptive if pushing a node into the stack removes the older version of the node, if it was present in the stack earlier.  We refer to the {\sf PUSH} operation in an adaptive stack as {\sf ADPPUSH} in the pseudocode. One way to implement an adaptive stack is via using a doubly linked list $L$ i.e., the algorithm stores the vertices in $L$ along with an array $P$ of $n$ pointers, one for each vertex, pointing to it's location in $L$. Now to push adaptively a vertex $v_i$, we first insert $v_i$ into $L$. Assuming it already belongs to $L$, go to $P[v_i]$ to update it so that it now points to the new location in $L$, and delete the older entry from $L$. Otherwise, $P[v_i]$ is empty, and is now updated to the newly inserted location of $v_i$. All of these can be done in $O(1)$ time. Popping a vertex $v_i$ is straightforward as we have to delete the node from $L$ and update $P[v_i]$ to {\sf NULL}. We also maintain in a bitmap of size $n$, call it {\it visited}, information regarding whether a vertex $v$ is visited or not. Then using all this auxiliary structure, it is easy to see that $BDS_{j}$ can be implemented in $O(m+n)$ time using $O(n)$ words or equivalently $O(n \lg n)$ bits of space (becuase of storing the list $L$ and the array $P$). This concludes a brief description of $BDS_j$ as well as its implementation. Jiang also showed, using $BDS_{j}$, how one can perform topological sort and strongly connected component decomposition. For detailed description, readers are referred to his paper~\cite{Jiang93}. Our focus here is to implement $BDS_{j}$ space efficiently. 

\begin{algorithm}
\caption{$BDS_{j}$($v$)}
\label{alg:bds}
\begin{algorithmic}[1]
\State {\sf EMPTY}(visited); {\sf EMPTY}(adp-stack);
\State {\sf ADPPUSH}($v$, adp-stack);
\While {{\sf ISNOEMPTY}(adp-stack)}
\State $w:=$ {\sf TOP}(adp-stack); 
\If {$w \notin $ visited}
\State visited := visited $\cup \{w\}$
\ForAll{$u$ in adj[$w$]}
\If {$u \notin $ visited}
\State {\sf ADPPUSH}($u$, adp-stack);
\EndIf
\EndFor
\Else \ {\sf POP}(adp-stack);
\EndIf
\EndWhile
\end{algorithmic}
\end{algorithm}

\begin{figure}[h]
\begin{center}
 \includegraphics[scale=.8, keepaspectratio=true]{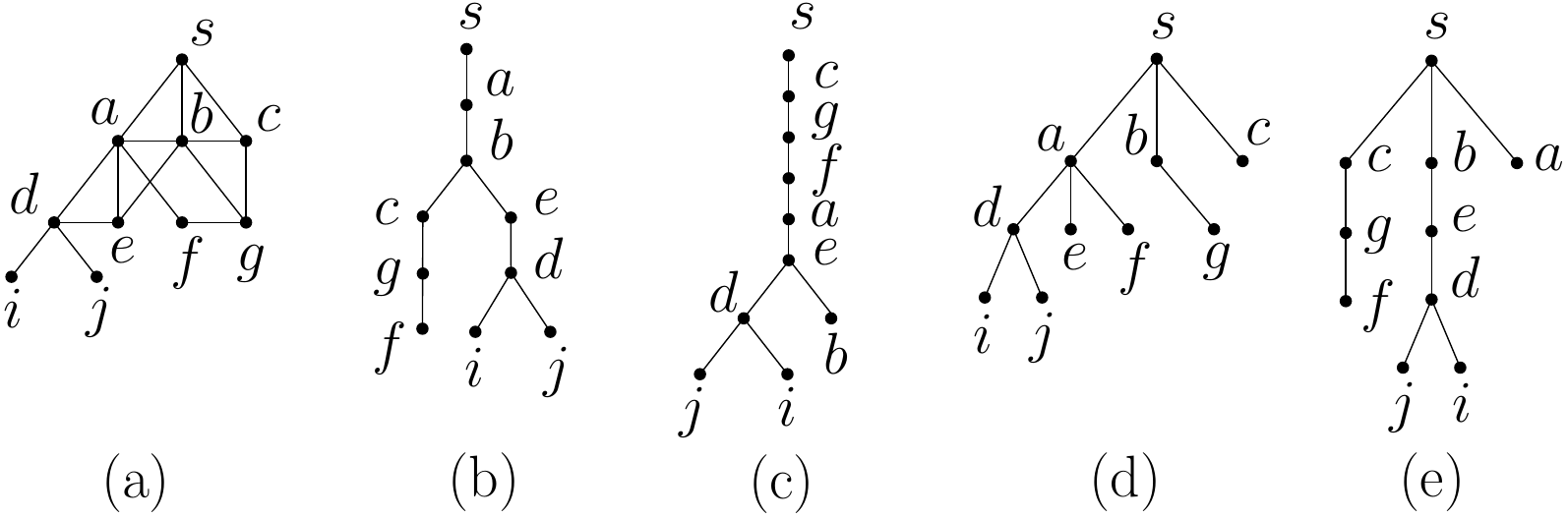}
\end{center}
\caption{(a) Graph $G$. We output the vertices when they are visited (for {\sf DFS} and {\sf BFS}) or expanded (for {\sf BDS}) for the first time in any graph search method. The adjacency lists are assumed to be ordered in the sorted order of their labels. (b) {\sf DFS} tree of $G$ and the resulting output for this {\sf DFS} traversal is $s,a,b,c,g,f,e,d,i,j$. (c) $BDS_j$ tree of $G$ and the resulting output for this $BDS_j$ traversal is $s,c,g,f,a,e,d,j,i,b$. (d) {\sf BFS} tree of $G$ and the resulting output for this {\sf BFS} traversal is $s,a,b,c,d,e,f,g,i,j$. (e) $BDS_{hs}$ tree of $G$ and the resulting output for this $BDS_{hs}$ traversal is $s,c,g,f,b,e,d,j,i,a$.}
\label{dfs_pic}
\end{figure}

In what follows, we illustrate a bit more on the inner working details of $BDS_{j}$ with the help of an example.  Following the convention, as in the recent papers~\cite{AsanoIKKOOSTU14,Banerjee2018}, here also in $BDS_{j}$ we output the vertices as and when they are expanded (note that, if reporting in any other order is required, it can be done so with straighforward modification in our algorithms). Hence the root will be output at the very first step, followed by its rightmost child and so on. Towards designing space efficient algorithms for $BDS_{j}$, we first note its similarities with {\sf DFS} traversal method. Taking the graph $G$ of Figure~\ref{dfs_pic}(a) as running example where (say) the root is $s$, and assuming that the adjacency list of every vertex is lexicographically sorted in the order of their labels, {\sf DFS} would have put $s$ first in the stack, followed by pushing $a$ then $d$ and so on. As a result, these three vertices would come first in the output of 
{\sf DFS} and so on. $BDS_{j}$ works in a slightly different manner. More specifically, $BDS_{j}$ pushes $a,b$ and $c$ into the stack (with $a$ at the bottom and $c$ at the top), and then expands $c$ (see the pseudocode for $BDS_{j}$). The node $b$ will again be discovered while expanding $c$, and due to the adaptivity of the stack, the older entry of $b$ which was inserted into the stack due to the expansion of $s$, will be removed (with a new entry of $b$ added to the stack). This phenomenon will be repeated again while expanding $g$. Eventually $b$ will be discovered from $e$ and expanded. See the final $BDS_{j}$ tree in Figure~\ref{dfs_pic}(c). To enable expanding a vertex during the last visit (instead of the first visit which is the case for {\sf BFS} and {\sf DFS}), Jiang used the adaptive stack. As analyzed previously, the bottleneck factor in the space consumption of $BDS_{j}$ is the adaptive stack. Our main observation is that we can get rid of the adaptive stack and still perform $BDS_{j}$ traversal of the graph $G$ correctly. More specifically, in what follows we describe how to implement $BDS_{j}$ space efficiently using a standard stack (without the adaptive push operation), along with some bookkeeping, yet producing the same vertex ordering as Jiang's $BDS_{j}$.

\subsection{Using $O(n)$ Bits and $O(m \lg n)$ Time}
\label{sec:oh-n-bits}
Note that, a vertex $v$ could be in one of the three states during the entire execution of the algorithm, (i) unvisited, (ii) expanded but not all of its children are expanded yet, and (iii) completed i.e., it is expanded as well as all of its children, if any. In our space efficient implementation of $BDS_{j}$, we denote them by color white, grey and black respectively, and store this information using an array (say) $C$ of size $O(n)$ bits. Along with this, we also store the last $O(n/\lg n)$ vertices that are grey in a (normal i.e., not adaptive) stack $S$. We divide the stack $S$ into blocks of size $O(n/\lg n)$ vertices where the first block refers to the first set of $O(n/\lg n)$ vertex labels that are pushed into $S$, the second block refers to the second bunch of $O(n/\lg n)$ vertex labels pushed into $S$ during $BDS_j$ and so on. Thus, there are $O(\lg n)$ blocks in total, and we always store the last block. Moreover, for every block we store the first and last element that are pushed in $S$ in a separate smaller sized stack $T$. Thus, we need overall $O(n)$ bits. 

Now armed with these data structures, we start by marking the root $s$ as grey and pushing in $S$. Note that, as and when a vertex $v$ gets expanded, i.e., turns grey, we can also output $v$ (i.e., report $v$ as the next vertex in $BDS_j$ order). At the next step, instead of inserting all of $s$'s white neighbors as in Jiang's $BDS_j$ implementation, we insert only the rightmost white neighbor $c$ into the stack, and change its color (from white) to grey (see Figure~\ref{dfs_pic}(c)). Observe crucially that by delaying the insertion of other white neighbors at once, we are essentially removing the need of adaptivity from the stack as now elements are pushed only when they are expanded, not multiple times as in $BDS_j$. Thus, we scan $c$'s adjacency list from right to left and insert the first white neighbor into the stack, mark it as grey in the $C$ array, and continue. We call this phase of the algorithm as {\it forward step} i.e., the phase in which we discover new vertices of the graph and insert them in $S$. At some point during the execution of the algorithm, when we arrive at a vertex $v$ such that none of $v$'s neighbors are white, then we color the vertex $v$ as black, and we pop it from the stack. If the stack is still non-empty, then the parent of $v$ (in the $BDS_j$ tree) would be at the top of the stack, and we continue the $BDS_j$ from this vertex. On the other hand, if the stack becomes empty after removing $v$, we need to reconstruct it to the state such that it holds the last $O(n/\lg n)$ grey vertices after all the pops done so far. We refer to the following phase of the algorithm as {\it reconstruction step}. For this, we basically repeat the same algorithm but with one twist which also enables us now to skip some of the vertices during this reconstruction phase. In detail, we again start with an empty stack, insert the root $s$ first and scan its adjacency list from the rightmost entry to skip all the black vertices and insert into the stack the rightmost grey vertex. Then the repeat the same for this newly inserted vertex into the stack until we reconstruct the last $O(n/\lg n)$ grey vertices. As we have stored the first and last vertices of each of the blocks in $T$, we know when to stop this reconstruction procedure. Another equivalent way to achieve the same effect is to recolor all the grey vertices back to white, while retaining the colors of all the other (black and white) vertices, and repeat the same algorithm. It is not hard to see that this procedure correctly reconstructs the latest set of grey vertices in the stack $S$. We continue this process until all the vertices become black. Obviously this procedure takes $O(n)$ bits of space. To bound the running time, note that, whenever this procedure tries to reconstruct, $O(n/\lg n)$ vertices have changed their colors to black, and they are not going to be inserted again into the stack. As this can happen only for $O(\lg n)$ rounds, and since 
in each round we might spend $O(m)$ time to scan the adjacency list and insert correct vertices into the stack, 
overall this procedure takes $O(m \lg n)$ time. We conclude this section by mentioning that a similar kind of idea was used in~\cite{AsanoIKKOOSTU14} to provide space efficient {\sf DFS} implementation, but we emphasize that ours algorithm is markedly different than~\cite{AsanoIKKOOSTU14} from the point of view of introducing {\it delayed insertion} of vertices into the stack, and thus removing the adaptivity from the stack, both the features not present in {\sf DFS}. In what follows, we describe an improved algorithm generalizing the ideas developed in this section.

\subsection{Using $O(n)$ Bits and $O(m \lg^* n)$ Time}
\label{sec:bds1-lglglg}

In this section we first describe an algorithm that uses $O(n \lg \lg \lg n)$ bits to performs $BDS_j$ in $O(m+n)$ time with high probability, and modify it later to get an even improved algorithm. 
To obtain this, we first divide the stack $S$ into $O(\lg n/ \lg \lg \lg n)$ blocks of size $n \lg \lg \lg n / \lg n$ vertices each. We group $(\lg n/ \lg \lg n)$ blocks into a super-block; thus there are $O(\lg \lg n/\lg \lg \lg n)$ super-blocks, each having $O(n \lg \lg \lg n/ \lg \lg n)$ vertices. For each vertex $v$, we store its 
(a)~color, 
(b)~super-block ID (SID), if it is in $S$, (and $-1$ if it is not added to $S$ yet, i.e., if it is white), and
(c)~the number of groups of $m/n$ vertices that have been explored with $v$ as the current vertex.
We also keep track of the first and the last element of each block, as well as super-block, and these takes up negligible (poly-logarithmic) space.  We describe the algorithm below in detail.

The algorithm is similar to the $BDS_j$ algorithm of Section~\ref{sec:oh-n-bits} with the following changes. The {\it forward step} remains mostly the same except updating the Items~(b) and~(c) above after every insertion of a vertex into the stack $S$. More specifically, whenever a vertex is inserted into the stack, we store its SID in an array (Item~(b) above), and also update the information regarding Item~(c) above (also stored in a separate array). In addition, we store the nodes in the topmost two blocks of the top super-block of the stack. We also maintain the block IDs (BIDs) of all the vertices belonging to the topmost two super-blocks using the dictionary structure of Theorem~\ref{dict}. 

The {\it reconstruction step} changes significantly as we cannot really afford $O(m)$ time for the reconstruction of each super-block (like in Section~\ref{sec:oh-n-bits}); rather we would ideally like to spend time proportional to the size of the super-block, hence resulting in an optimal linear time algorithm. In order to achieve this, we do the following. As we have stored the first element of all the super-blocks, we can start by pushing that element (say $v$) into a temporary stack. We obtain the next vertex by determining (by consulting Item~(c) above) the first grey vertex, say $u$, that belongs to this super-block (as we can check from its SID) from the right endpoint of $v$'s adjacency array, and that is not already inserted in the current reconstruction procedure (can be checked from the dictionary structure of Theorem~\ref{dict}). Now we repeat the same in $u$'s list until we reconstruct the whole super-block. Note that, simultaneously we are also inserting the BIDs for every vertex in the structure of Theorem~\ref{dict}. We should mention one point at this time, the necessity of dynamic dictionary comes from the fact that we need to quickly find the BID information associated with the vertices in order to decide whether to insert any particular vertex in the stack or not. For performing this task very efficiently both time and space wise, having a simple array is not enough and thus, the requirement of more powerful dynamic dictionary structure. Due to the space limitations, we may need to discard all other blocks inside a super-block except the topmost two. Once we reconstruct the required blocks, the algoithm can proceed normally. Now all that is left is to determine the time and the space complexity of this procedure. Space requirement of our algorithm is $O(n \lg \lg \lg n)$ bits which is dominated by the SID, topmost two blocks inside the top super-block and the dictionary~structure.

To bound the number of reconstructions, note that, each time we reconstruct a super-block, the previous super-block's $O(n \lg \lg \lg n/ \lg \lg n)$ vertices change their color to black and get popped from the stack, hence they will never be pushed again. Thus, the number of restorations (denoted by $q$) is bounded by $O(\lg \lg n/\lg \lg \lg n)$. Now if the degree of a vertex $v$ is $v_d$, then we spend $O(\textit{min}\{v_d,m/n\})$ time on $v$ searching for its correct neighbor in our algorithm due to the information stored in Item~(c) above. To bound the running time of the algorithm, note that over $q$ reconstructions and over all vertices of degree at most $m/nq$, we spend $O(qn(m/nq))=O(m)$ time, and for vertices having degree larger than $m/nq$, over $q$ such reconstructions, we spend $O(q(n/q)(m/n))=O(m)$ time. Observe that, this running time is randomized linear because of the use of dynamic dictionary\footnote{Our algorithm performs atmost $O(m+n)$ insertion/deletion/retrieval during its entire execution using the dictionary of Theorem~\ref{dict} which takes $O(1)$ time with a probability of $(1-1/n^c)$ (where $c \geq 3$) for each insertion/deletion/retrieval. Thus, the probability that our algorithm takes more than $O(m+n)$ time is $(1/n^{c-2})$ by union bound rule.} of Theorem~\ref{dict}. This concludes the description of the $BDS_j$ algorithm taking randomized $O(m+n)$ time and using $O(n \lg \lg \lg n)$ bits with high probabality $(1-1/n^c)$ for some constant $c$. 

Before generalizing this algorithm, let us define some notations that are going to be used in what follows. The function $\lg^{(k)}n$ is defined as applying the logarithm function on $n$ repeatedly $k$ times i.e., $\lg \lg \ldots \textit{($k$ times)} \ldots \lg n$. Similarly $\lg^* n$ (also known as {\it iterated logarithm}) is the number of times the logarithm function is iteratively applied till the result is less than or equal to $2$. It's easy to see that $\lg^{(\lg^* n)}n$ is always a constant for any $n$. Like the previous algorithm, this algorithm also uses the data structures of Item~(a), (b), and (c) along with a hierarchy of levels (instead of just two levels like the previous algorithm). For some $k$ (which we will fix later), we set the size of $k$-th level blocks as $O(n/({\lg^{(k)}n})^2)$, and we divide the $k$-th level blocks into $(k+1)$-th level blocks. Thus, the number of $k$-th level blocks inside a $(k+1)$-th level block is $O(\{(\lg^{(k)}n)/(\lg^{(k+1)}n)\}^2)$, where $k=1$ means the smallest level blocks. We store the dynamic dictionary for $k$-th level at the $(k+1)$-th level for every $k$, and the space required for storing the dictionary at level $k$ is given by $O((n/({\lg^{(k+1)}n})^2)(\lg\{(\lg^{(k)}n)/(\lg^{(k+1)}n)\}^2))=o(n)$ bits. Through the entire execution of the algorithm, we always maintain the topmost two smallest level blocks along with other data structures. The {\it forward step} as well as the {\it reconstruction step} of the algorithm remains exactly the same other than modifying/storing informations at each level of the data structures suitably. As the work involved at each such level is simply one of the four operations from $\{$insertion/deletion/membership/retrieval$\}$ (all takes $O(1)$ time with high probability) at the dynamic dictionaries of the corresponding levels, by similar analysis as before, the final running time of the algorithm simply becomes $m$ times the overall number of levels of data structure that we maintain during the execution of the algorithm, and this can be bounded by $O(mk)$. Also, we can bound the overall space requirement as $O(n \lg^{(k+1)}n)+o(n)$ bits. Now choosing $k+1=\lg^*n$, our algorithm takes $O(n)$ bits of space and $O(m \lg ^*n)$ running time, and this concludes the description of the algorithm.

In what follows, we specially focus on designing space efficient algorithms for $BDS_j$ when the input graph is sparse (i.e., $m=O(n)$). Studying such graphs is very important not only from theoretical perspective but also from practical point of view. These graphs appear very frequently in most of the realistic network scenario, like Road networks and the Internet, in real world applications.

\subsection{Using $O(n \lg (m/n))$ Bits and $O(m+n)$ Time}
\label{sec:bds1onln}
In this section, we show how one can obtain linear bits and linear time algorithm for $BDS_j$ for sparse graphs. For this we use the following lemma from~\cite{Banerjee2018}.

\begin{lemma}\label{lem:adjlist-pointers}(\cite{Banerjee2018})
Given the adjacency array representation of a graph $G$,
using $O(m)$ time, one can construct an auxiliary structure of size $O(n \lg (m/n))$ bits that can store a ``pointer'' into an arbitrary position within the adjacency array of each vertex. Also, updating any of these pointers takes $O(1)$ time.
\end{lemma}

The idea is to store {\it parent} pointers into the adjacency array of each vertex using the representation of Lemma~\ref{lem:adjlist-pointers}. More specifically, for an undirected graph, whenever the $BDS_j$ expands a vertex $u$ to reach $v$ following the edge $(u,v)$, $u$ becomes the parent of $v$ in the $BDS_j$ tree, and at that time, we scan the adjacency array of $v$ to find $u$ and store a pointer to that position (within the adjacency array of $v$). For every vertex $v$ in $G$, we can also store another pointer marking how far in $v$'s list $BDS_j$ has already processed. This pointer will start from the very end of every list, gradually moves towards the left, and at the end of the algorithm, will point to the first vertex of list. We also maintain color information in a bitmap of size $O(n)$ bits. Given this pointer representation, it is easy to see how to implement $BDS_j$ in $O(m+n)$ time. The main advantage of this algorithm of ours is, note that, we don't even need to maintain any explicit stack to implement this process. We can extend similar idea for doing $BDS_j$ in directed graphs by setting up parent pointers (which are used during backtracking) in the in-adjacency list of every vertex and use the other pointer to mark progress in the out-adjacency list. With this, we complete the proof of Theorem~\ref{combined} for $BDS_j$.

\section{Breadth-depth Search of Horowitz and Sahni} \label{BDS2}
This version of {\sf BDS} works as follows. The algorithm starts by pushing the root (i.e., the starting vertex) into the stack $S$ initially. At every subsequent step, the algorithm pops the topmost vertex $v$ of $S$, and pushes all its unvisited neighbors into $S$.
See Figure~\ref{dfs_pic}(e) for an example.
Note crucially that, due to the popping of the parent while pushing the children in $S$, during backtracking the next vertex to be expanded is always at the top of the stack $S$. 
This stack $S$ could grow to contain $O(n)$ vertices, thus the classical implementation of this procedure takes $O(m+n)$ time and $O(n \lg n)$ bits of space. See~\cite{HorowitzS78} for a detailed description. In what follows, we show how to implement this $BDS_{hs}$ space efficiently.

\subsection{Using $O(n)$ Bits and $O(m \lg ^* n)$ Time}
To implement $BDS_{hs}$ using $O(n)$ bits, we crucially change the way we handle the stack during the execution of the algorithm. More specifically, we will not pop immediately the vertex $v$ which is going to be expanded at the very next step (as done in~\cite{HorowitzS78}), rather keep it in the stack $S$ instead for later use. We refer to this technique as the {\it delayed removal} of the vertices. Even though this is different than the {\it delayed insertion} technique (which was crucially used for $BDS_j$'s implementation), it is worth emphasizing that by introducing delayed removal of the vertices, the behaviour/operation of the stack in $BDS_{hs}$ becomes pretty similar to the one in $BDS_j$ (as it will be clear from the next paragraph), thus we can reuse previously developed ideas for $BDS_j$ to obtain space efficient implementation of $BDS_{hs}$.
In addition to this change, we use three colors as we did in the previous $BDS_j$ implementation with the exact same meaning attached to them, and store this information in an array $C$. Also, we always store the last block of $O(n/\lg n)$ grey vertices of $S$. 

In detail, we start by marking the root, say $s$, as visited, coloring it grey and inserting it into $S$. This is followed by inserting all of $s$'s unvisited white neighbors into $S$, change them to grey in $C$. Now $s$'s rightmost child (say $v$) is at the top of the stack and we insert in $S$ all of $v$'s white neighbors without popping $v$, also simultaneously marking them visited, and coloring $v$ as grey. This process is repeated until we arrive at the vertex $u$ all of whose neighbors are visited; at this point we make $u$ to be black and pop it from the stack. The vertex which is below $u$ in the stack (say $p$) is either its parent (if $u$ is the first child of its parent) or its previous sibling. We actually don't know which case it is, but it does not matter -- we simply continue the search from $p$. The case when $p$ is the previous sibling of $u$ is handled the same way by the original algorithm as well as ours.
In the case when $p$ is the parent of $u$, all the other children of $p$ are colored black (since $u$ is the first child of $p$), and hence our algorithm colors $p$ as black and pops it from $S$.
Reconstructions are also handled in a similar fashion as in Section~\ref{sec:oh-n-bits}. I.e., we recolor the grey vertices back to white, and start executing the same algorithm from root but we don't insert the black vertices again. This ensures that, if a vertex has become black already, its subtree will not be explored again, and once we restore the latest block of $O(n/ \lg n)$ vertices, we start executing the normal algorithm. Clearly, we are using $O(n)$ bits of space. Since the reconstruction happens only $O(\lg n)$ times, and each time we spend $O(m)$ time, overall this procedure takes $O(m \lg n)$ time. Generalizing this strategy by creating hierarchy of levels and then using dynamic dictionary at each levels like we did for $BDS_j$ in Section~\ref{sec:bds1-lglglg}, we can similarly obtain an implementation of $BDS_{hs}$ taking $O(n)$ bits and $O(m \lg ^* n)$ time. This completes the description of the algorithms taking $O(n)$ bits.

\subsection{Using $O(n \lg (m/n))$ Bits and $O(m+n)$ Time}
We can use Lemma~\ref{lem:adjlist-pointers} to store parent pointers in the adjacency array of every vertex, and another pointer to mark the progress of $BDS_{hs}$ so far in a similar way as we did for $BDS_j$ in Section~\ref{sec:bds1onln}. It is easy to see that with these structures, and additional color array, using $O(m+n)$ time and $n \lg (m/n)$ bits, we can implement $BDS_{hs}$. One can also extend this to the directed graphs as metioned in Section~\ref{sec:bds1onln}. With this, we complete the proof of the Theorem~\ref{combined} for $BDS_{hs}$.

\section{Conclusions}\label{conclusion}
We obtained space-efficient as well as time-efficient implementations for two graph search methods, both are known under the same name, breadth-depth search even though they perform entirely differently. The main idea behind our algorithm is the  introduction of the {\it delayed insertion} and the {\it delayed removal} techniques for better managing the elements of the stack, and finally we use the classical blocking idea carefully to obtain the space-time efficient implementations. We think that these ideas might be of independent interest while designing similar space-time efficient algorithms for other existing graph search methods in the literature. We believe this is an important research direction as these search methods form basis of many important graph/AI algorithms. 

We leave with two concrete open problems, is it possible to design a) $o(n)$ space and polynomial time algorithms, and b) $O(n)$ bits and $O(m+n)$ time algorithms (deterministic or randomized) for both the {\sf BDS} implementations? Another interesting direction would be to study these graph search methods in the recently introduced in-place~\cite{Chakraborty00S18} model where changing the input is also allowed in a restricted manner unlike the {\sf ROM} model which is what we have focused in this paper.

\bibliographystyle{plain}
\bibliography{dfs}

\end{document}